\documentstyle[12pt,epsf,twoside,fleqn,espcrc1]{article}
\begin{document}

% put your own definitions here:
\newcommand{\ipb}{$^{208}$Pb~}
\newcommand{\ttbs}{\char'134}
\newcommand{\AmS}{{\protect\the\textfont2
  A\kern-.1667em\lower.5ex\hbox{M}\kern-.125emS}}

% add words to TeX's hyphenation exception list
\hyphenation{author another created financial paper re-commend-ed}

% declarations for front matter

\title{
Barrier Distributions as  a Tool to Investigate Fusion and Fission
}
\author{M.~Dasgupta,\address{Department of Nuclear Physics, Research School of 
Physical
Sciences and Engineering,\\ The Australian National University,
Canberra ACT 0200, Australia\\}
D.J.~Hinde$^{a}$, J.R.~Leigh$^{a}$ and 
K.~Hagino\address{ Department of Physics, Tohoku University, Sendai 980--77,
Japan}}

\maketitle

\begin{abstract}
The recent availability of precisely measured 
fusion cross-sections has enabled the extraction of a representation of 
the distribution of 
barriers 
encountered during fusion. These representations,
obtained from a variety of reactions, 
provide a direct observation of {\it how} the structure of the fusing nuclei
changes the inter--nuclear potential landscape, thus affecting
the fusion probability. Recent experiments showing the effects of 
static quadrupole and hexadecapole 
deformation, single-- and double--phonon states, transfer of nucleons 
between two nuclei, and  high lying excited states are reviewed.
The application of these concepts to the explanation of the 
anomalous fission--fragment  anisotropies observed following
reactions with actinides is discussed. 
\end{abstract}

\section{INTRODUCTION}

In the simplest fusion model, two nuclei are
assumed to fuse 
once they have penetrated the one-dimensional 
real potential barrier resulting from the sum of the 
repulsive Coulomb and centrifugal 
potentials, and the attractive, short range nuclear potential. 
However, for reactions involving 
heavy ions, the measured fusion cross-sections 
were found to be significantly different from
the expectations of such a model, particularly at energies below 
the barrier where enhancement of several orders of magnitude was
observed~\cite{st86,be88}. 
The inadequacy of this model was
 conclusively demonstrated~\cite{ba83} by inverting the experimental data, which 
yielded unphysical inter--nuclear potentials.  It was shown 
subsequently~\cite{da82} that the 
observed  enhancements were due to coupling of the translational motion to 
intrinsic degrees of freedom, {\it e.g.} rotation, or surface vibrations, 
 a fact which is not taken into account in the single-barrier 
penetration model.

The role of static deformation effects~\cite{wong,vaz,stok} in enhancing
sub-barrier fusion can easily be visualised in a classical picture, where 
different orientations of the deformed target nuclei are encountered during 
the collision, giving rise to a distribution of barrier heights, some lower 
and some higher than the single-barrier. 
Due to the exponential dependence of the fusion probability on the barrier 
height, any distribution of barriers which gives a barrier at an energy 
lower than the 
single-barrier, $B_0$, leads to enhancement of the cross-sections
at energies below $B_0$.
In general the effect of couplings to other degrees of 
freedom results in a change from a single-barrier penetration scenario 
to  multi--dimensional barrier penetration.
In the next section, the concept of the distribution of barriers
is discussed. The experimental results, spanning strongly-coupled  to 
weakly-coupled systems, and their interpretations, are presented 
in Secs.~3--6. The correlations and applications 
of these new results to other fields are discussed in Sec.~7, followed by a
summary in Sec.~8.

\section{FUSION BARRIER DISTRIBUTIONS}

The above discussion indicates that viewing the collision process as two  
structureless spheres interacting only through the inter--nuclear potential 
is too simplistic, and coupling to other reaction channels has to be 
considered. Theoretical calculations including coupling to inelastic
and transfer channels are usually carried out by solving a set of 
coupled--channels equations. The effects of such couplings, however can 
easily be visualised in the 
eigenchannel approximation, where it was shown~\cite{da82} that the effects of 
such couplings 
are to replace the single-barrier by a distribution of barrier heights, thus 
providing an explanation of the observed disagreement~\cite{st86,be88} with the 
single-barrier penetration calculations. This improved understanding 
was pivotal for interpreting the experimental results. However 
since any coupling mechanism would give rise to
enhancement, an educated guess was required for the type of coupling 
to be included in a given system.

This point is 
illustrated in Fig.~\ref{fig1}, where three theoretical excitation functions 
 for reactions involving different coupling schemes are shown. 
The  coupling schemes  involve coupling to a negative
$Q$-value channel in Fig.~\ref{fig1}(a),  a positive $Q$-value channel 
in Fig.~\ref{fig1}(b),
and coupling associated with a deformed nucleus in Fig.~\ref{fig1}(c).
The discrete barriers associated (under the eigenchannel approximation) 
 with the first two cases 
are shown in Fig.~\ref{fig1}(d) and (e) by the thick vertical lines with 
lengths indicative of the probability of encountering the barrier;  
the third coupling scheme will give a continuous 
distribution of barrier heights associated with different orientations of the
deformed nucleus. 
Comparison 
with the dashed lines resulting from a single-barrier penetration 
calculation, shows that the enhancements at the lower energies 
are very similar in the 
three 
cases, despite the differences in the nature of couplings. 
There are significant differences in the  excitation functions at other 
energies but the type of 
coupling involved is not immediately apparent from the excitation functions
themselves.

A few years ago it was suggested~\cite{ro91} 
that a representation of the distribution of barrier
heights in a reaction could be extracted {\it directly} from a fusion
excitation function.
In a classical sharp cut-off model,
\begin{equation}
 E\sigma =
{\sum_\alpha}w_{\alpha}\pi R^2_{\alpha}(E-B_{\alpha}),
\end{equation}
 where $E$ is the energy, $\sigma$ is the fusion cross--section, 
$R_{\alpha}$ is the fusion radius,  $w_{\alpha}$ is the
probability of encountering the barrier 
and $B_{\alpha}$ is the barrier energy associated with
the channel index $\alpha$.
In this case, $d^2(E\sigma)/dE^2$ returns the
original discrete barrier distribution as a
set of delta functions, centered at the barrier energies and  
 normalised by the $w_{\alpha}$~\cite{ro91}.
When quantum mechanical penetration of the barriers is considered, the
cross-sections vary smoothly in the vicinity of each
barrier and $d^2(E\sigma)/dE^2$ becomes continuous,
replacing the delta functions by a near Gaussian function~\cite{ro91}. 
It was also shown~\cite{le95} that the function $d^2(E\sigma)/dE^2$ 
does indeed correspond closely to the ``true'' barrier 
distribution smoothed by the quantum tunnelling effects.

Applying this technique to the excitation functions of Fig.~\ref{fig1}
returns a function shown by the solid lines in the lower panels. It can be
seen that they are very different for the three cases, 
and reveal directly where the barrier strength lies. For example, 
in Fig.~\ref{fig1}(d) and (e) the solid curves are
 centered around the calculated discrete 
barrier positions (thick lines) and the heights of the gaussians reflect the 
probability of encountering the barrier. In the third case the solid curve 
coincides with the barrier distribution expected for a deformed nucleus.   
It should be noted that the interpretation of an experimental
barrier distribution is not always as  simple as suggested by
the examples shown in
Fig.~\ref{fig1}. These are
idealised cases and in reality several different types of coupling may be
involved in a given reaction, complicating the situation. Nonetheless, the
ability to see directly how barrier strength is distributed is useful
 in understanding the reaction process.

The function 
$d^2(E\sigma)/dE^2$ will be referred to as 
the barrier distribution  in this contribution, even though it is 
recognised that it gives only a representation of the 
true barrier distribution.
Also, since the experimental data are measured at discrete energies, the 
function 
 $d^2(E\sigma)/dE^2$ 
is extracted using a point difference formula. For consistency,
when comparing an experimental distribution with theory,
the theoretical distribution is evaluated in an
identical  manner.

\section{TESTING THE CONCEPT - REACTIONS INVOLVING ROTATIONAL NUCLEI}

The concept of a fusion barrier distribution outlined above is particularly 
attractive,  
since a representation 
can be directly extracted from experimental data, 
thus reducing the uncertainty in the type of coupling schemes to be 
introduced to explain the experimental results. Fig.~\ref{fig2}(a)
 shows the barrier 
distribution for the $^{16}$O~+~$^{154}$Sm reaction, 
extracted from data~\cite{stok} which has an experimental 
uncertainty of 10\%; a level 
typical for the data prior to the new generation of high 
precision measurements. Such data
yield poorly defined barrier distributions and can therefore
be equally well
reproduced with models incorporating very different barrier
distributions, as shown by the dashed and solid lines.  
Thus testing the concept of barrier 
distributions, which requires the estimation of the second 
derivative of the experimentally measured quantity $\sigma E$  with 
respect to E, needs the fusion excitation 
function to be measured to much higher levels of precision. 

This necessity prompted experiments~\cite{we91} at the Australian National 
University, where
high precision (uncertainty $\sim$1\%) measurements of 
fusion excitation functions were made.
The barrier distribution for the $^{16}$O~+~$^{154}$Sm reaction, 
extracted from these data is shown in Fig.~\ref{fig2}(b). The advantages of 
high precision measurements are clear in defining the  
experimental distribution at the level where it can now distinguish between 
the two different calculations. 
The reactions of  $^{16}$O with deformed $^{154}$Sm and $^{186}$W nuclei 
were chosen to serve
as an experimental test of the barrier distribution concept, since the
effects of
deformation are well established, and the barrier distributions were
expected to be close to the classical ones, which are readily calculable.

The measured~\cite{we91,le93} excitation functions for $^{16}$O~+~$^{154}$Sm and 
$^{186}$W  are shown in the upper panels of Fig.~\ref{deform}. 
The details of the experimental set--up used to obtain these data 
and the precautions taken to obtain the  low experimental 
uncertainties are described 
in Ref.~\cite{le95}. The barrier distributions extracted 
for  these two systems are 
very different as shown in the 
the lower panels of Fig.~\ref{deform}.
 The dotted curves are the results of
calculations using  a single-barrier penetration model.
The obvious 
features are that the experimental cross-sections are underestimated and 
the experimental distributions are very wide relative to 
the calculations using a single-barrier. 
The inclusion of the effects
of the static quadrupole deformations of the $^{154}$Sm and $^{186}$W 
nuclei in the calculations (dashed line) dramatically improves the
agreement with experiment, producing  a wide 
barrier distribution, arising from the random
orientations of the deformed nuclei.
In fact comparing the calculated and experimental excitation functions, 
 as done traditionally, one could be led to believe that they agree 
very well. The barrier distributions however give a different picture; 
the calculations disagree with the measurements at E$\sim$56 MeV and 66 MeV 
for $^{154}$Sm and $^{186}$W respectively and they do so in 
different ways for the two reactions. This  suggests that the data 
are sensitive 
to nuclear properties  other than the quadrupole deformation. 

The $^{154}$Sm and $^{186}$W nuclei are known to have significant 
hexadecapole deformations which are similar in magnitude but 
differ in sign; the former has a positive $\beta_4$ and the latter a 
negative one. The results of including the effects of 
hexadecapole deformations in the calculations
are shown by the solid lines, which agree remarkably well with the 
experimental results. It is clear that the differences in the shape of the 
calculated (and experimental)  barrier distributions for the two systems 
is mainly due to the differences in the sign of the $\beta_4$ value. 
Some minor discrepancies between these calculated and the 
experimental barrier distributions which are still present, are 
 discussed in Ref.~\cite{le95}. 

The results of these  measurements exceeded all expectations, showing 
that fusion excitation functions are not only very sensitive to 
quadrupole deformations but also to small 
hexadecapole deformations as well. These measurements also made it possible 
to apply the proposition of extracting fusion barrier distribution from the 
experimental data, and showed  that for permanently deformed nuclei 
the function $d^2(E\sigma)/ dE^2$ does indeed return a 
sensible representation of barriers {\it expected} to be 
encountered during the reaction.

\section{VIBRATIONAL NUCLEI -- DOES THE CONCEPT STILL WORK ?}

The barrier distributions extracted for the rotational nuclei provided a 
good test of this new approach to fusion. Since for these cases the 
concept of a ``distribution'' had a classical analogue, it was 
crucial to test whether this picture would still be applicable in 
cases where the existence
of a distribution of barriers is not classically  apparent. Further, the concept of 
a distribution is strictly valid for cases with zero excitation energy
(eigenchannel approximation) and thus it was not clear whether couplings to 
vibrational states with typical 
excitation energies of a few MeV will give rise to the simple barrier 
distribution as predicted in Ref.~\cite{da82}. The reactions 
$^{16,17}$O~+~$^{144}$Sm were chosen for this purpose as 
the coupling scheme is expected to be relatively simple,
because the degree of collectivity associated with the semi-magic nucleus
$^{144}$Sm is relatively small. The use of $^{17}$O was to check if the 
distributions are sensitive to the additional neutron in the projectile.

 The measured~\cite{mo94} excitation functions along with the extracted 
barrier distributions for the $^{16,17}$O~+~$^{144}$Sm 
 reactions are shown in the 
upper and lower panels of Fig.~\ref{sm144}.
Calculations assuming no
coupling are shown by the dotted lines; clearly they
are inconsistent with experiment.
The experimental barrier distributions for both
reactions are dominated by a two--peaked structure, with 
the  main strength around 60 MeV and a peak with smaller strength 
near 65 MeV; this picture closely resembles that of Fig.~\ref{fig1}(d) , a 
typical case of coupling to --Q value channels.
Since the two
reactions have significantly different $Q$-values for one- and two-particle
transfer reactions, and the projectiles have different structure, the
consistency of this
feature strongly implies that it is associated with inelastic excitation
of $^{144}$Sm. The most important channels associated with excitation
of $^{144}$Sm are expected to involve those states with the
largest B(E$\lambda$)$\uparrow$ values, {\it i.e.} the first 2$^+$ and 3$^-$ 
states. Considering firstly the $^{16}$O reaction, including the
2$^+$ and 3$^-$ states of  $^{144}$Sm in the calculation gives a good 
representation~\cite{le95}
of the data, as shown in Fig.~\ref{sm144}(a) and (b).

In the reaction with the $^{17}$O projectile, 
the excitation function and the barrier distribution at higher energies 
are very similar to those of the reaction with $^{16}$O, but 
at low energies the cross-sections for the $^{17}$O reaction are
more than four times higher than those for $^{16}$O.
A calculation including couplings to the 
2$^+$ and 3$^-$ states of $^{144}$Sm, similar to that which describes the 
$^{16}$O reaction very well, shown by the 
dashed line in Fig.~\ref{sm144}(c) and (d), 
fails to fit the low energy data. It is clear 
from the barrier distribution that this calculation gives too much 
strength around the main peak and fails to reproduce the observed 
low energy tail, indicating a need for additional coupling.
An increased cross-section will result from any form of
extra coupling, as was shown in Fig.~\ref{fig1}, but  the type of 
coupling required is not immediately apparent in the excitation function.
However, referring to Fig.~\ref{fig1}(e), a low energy tail in the barrier 
distribution 
would imply couplings to a +Q-value channel, which rules  out couplings to 
further inelastic channels, and indicates couplings to transfer channels.  
The neutron stripping 
reaction $^{144}$Sm($^{17}$O,$^{16}$O)$^{145}$Sm  has a ground-state 
$Q$-value of $+$2.6 MeV and is the obvious candidate for inclusion in the
calculations. The calculations including couplings to this transfer channel 
give an extremely good representation of the data~\cite{le95}
 as shown by the solid 
line in the right panels of Fig.~\ref{sm144}.

The excellent agreement
between the theoretical calculations and the experimental results strongly
supports the  barrier distribution picture, even when the excitation 
energies are not very close to zero as in the eigenchannel picture,
 thus indicating that this technique can 
potentially be used as a tool to give  a better understanding of 
more complex reaction processes. These experiments also  
unambiguously demonstrate the effects on fusion of couplings to 
inelastic 
and transfer channels. Since these experiments, 
measurements for the reactions $^{32,36}$S~+~$^{110}$Pd and  
$^{40}$Ca~+~$^{46,48,50}$Ti have been studied at other 
laboratories~\cite{stepd,cati} 
and have nicely demonstrated the effects of couplings to transfer 
channels on the barrier distributions.

\section{UNRAVELLING COMPLEX PHONON COUPLINGS}

The successful interpretation of the fusion excitation function for the 
above--mentioned systems and consequently the clear identification of the 
couplings affecting fusion prompted further experiments to see and/or 
 identify how more exotic forms of couplings affect the fusion process.
A study~\cite{st95} of the $^{58}$Ni~+~$^{60}$Ni reaction, described below, 
provided a surprising and 
striking result due to the effects of complex surface vibrations in fusion.
Experimental studies~\cite{mo95} of the fusion of two doubly closed shell nuclei,
the $^{16}$O + \ipb system, yielded yet another surprising result which is 
discussed in Sec.~5.2.

\subsection{The $^{58}$Ni~+~$^{60}$Ni reaction}

This system was studied~\cite{st95} at the Laboratori Nazionali di 
Legnaro, Italy, in
order to address the question of whether sub--barrier fusion enhancement is mainly due to 
an  elastic ($Q$ = 0) two neutron transfer channel, or due to couplings to 
vibrational states in the two Ni isotopes. The results have been discussed 
in detail by N. Rowley in this conference and hence here we quote the main 
results only. It was found that the barrier distribution had a 
three-peaked structure, a feature which could not be reproduced by 
coupling to neutron--pair transfer and/or coupling to simple 1--phonon 
states in the Ni nuclei. It was shown~\cite{st95} that couplings not only to the
1--phonon states  but more importantly to 2--phonon states in each nucleus
(and their mutual excitations) were necessary to obtain a good 
representation of the data. This was the first time the effects of such 
complex couplings were clearly identified as affecting the fusion 
process.

\subsection{The $^{16}$O~+~$^{208}$Pb reaction}

It might be believed {\it a priori} that the results of fusion  of two doubly 
closed shell nuclei should be easily explained. Interestingly, 
previous~\cite{opbdiff} 
experiments and analyses encountered difficulties in explaining 
the fission fragment anisotropies and the deduced 
fusion mean square angular momentum.
High precision fusion cross-section 
measurements were made~\cite{mo95} with the aim of understanding some of these features.
The measured  fusion excitation function
and the barrier distribution are presented in Fig.~\ref{opb}.
The single-barrier penetration calculations (dashed lines) 
underpredict the excitation function, and 
fail to reproduce the wide barrier distribution,
clearly indicating the presence of couplings with  other channels.

The solid lines in the left panels of Fig.~\ref{opb} show the 
 results of a realistic 
coupled--channels calculation~\cite{ha97} which includes couplings to 
the 2$^+$, 3$^-$ and 5$^-$ vibrational states in \ipb (discussions 
regarding the non-inclusion of transfer channels is detailed in 
Ref.~\cite{da97}). 
The calculations 
predict a double-peaked structure as opposed to the more complex structure 
seen in the experimental barrier distribution, and also underpredict 
the low energy part of the 
excitation function. 
It can easily be shown~\cite{ro93} 
that the double-peaked structure of the calculated barrier 
distribution will remain essentially unchanged even when couplings to 
other 1--phonon states are considered. Neither can the agreement 
be improved by increasing the coupling strength.
 It is thus clear that couplings from the elastic channel to 
single--phonon states in $^{208}$Pb are not sufficient to explain the data.

Using an eigenchannel approximation
it has been  shown~\cite{ro93} that 
the introduction of couplings to 2--phonon states results in three barriers, 
with the two  lower barriers being close to each other and the third barrier 
lying at higher energy. 
In the present case, the 
experimental barrier distribution would seem to indicate this scenario.
The existence of 2--phonon octupole excitations in \ipb was recently 
shown~\cite{ye96} experimentally. Results of the coupled--channels 
calculations, 
including the 2$^+$, 3$^-$, 3$^- \otimes$3$^-$ and 5$^-$ vibrational states 
in $^{208}$Pb,  and all the resulting cross--coupling terms 
{\it e.g.}, 2$^+ \otimes$3$^-$ 
{\it etc.}, are shown by the solid lines in the right panels of 
Fig.~\ref{opb}.
It is clear that the agreement with the 
experimental  excitation function and the shape of the barrier distribution 
is dramatically improved~\cite{da97} with the inclusion of couplings to 2--phonon 
states.  

These experimental  results thus 
 support the recent spectroscopic measurements 
indicating the presence of a double--phonon state in 
$^{208}$Pb. It is interesting that the dynamics of the fusion process even 
for reactions
between two closed shell nuclei, which might be thought to be  simple,
particularly at low energies, is affected by complex surface
vibrations.

\section{ROLE OF CLOSED SHELL NUCLEI}

As seen in the previous sections, a successful description of the fusion
cross-sections for reactions involving $^{16}$O did not require 
couplings to states in $^{16}$O. This is particularly curious as 
the octupole vibrational state has a large strength 
and indeed previous work~\cite{oexcit,oexcit2} on the $^{16}$O + $^{154}$Sm, $^{A}$Ge reactions 
included couplings to this state.
In marked contrast, no specific features appear
in the measured barrier distribution for the $^{16}$O + $^{144}$Sm
reaction which can be associated with the excitation
of $^{16}$O; rather it was seen~\cite{le95} that a good
theoretical representation of the experimental fusion
barrier distribution is destroyed when the projectile excitation is 
included. 
Recent measurements of $^{40}$Ca~+~$^{194}$Pt, $^{192}$Os~\cite{bi96}
at the University of Washington, Seattle, 
show a characteristic structure with a high energy peak arising due to 
coupling to the octupole state of $^{40}$Ca, as indicated by the vertical 
arrows in Fig.~\ref{caos}.

Both of these conclusions are based on the comparison of the experimental
data with the results of simplified coupled--channels 
calculations~\cite{ni89,da92}, where
the linear coupling approximation is used to describe the vibrational
excitation of the projectile. The inadequacy of the 
linear coupling approximation
has been pointed out by several authors~\cite{ha97,esquad}, and it is 
likely that this  approximation does not give the true fusion barrier 
distribution for 
couplings to the 3$^-$ state in $^{16}$O,  which has a large E3 
strength.
The results of realistic coupled--channels calculation~\cite{ha97} under 
the linear coupling approximation,  for 
the $^{16}$O~+~$^{144}$Sm reaction with and without 
couplings to the 3$^-$ state in $^{16}$O 
are shown in the left panels of 
Fig.~\ref{allord}. 
It is seen that couplings to the 3$^-$ state in $^{16}$O gives 
essentially a single--peaked barrier distribution and  destroys the good 
agreement with the data obtained by ignoring it, as also observed 
in Ref.~\cite{le95}.

Performing similar realistic calculations, where however 
the couplings to the octupole vibrations of both $^{16}$O and $^{144}$Sm
are treated to all orders, gives the result~\cite{prl97} shown
 in the right hand panels of Fig.~\ref{allord}.
It is remarkable that these calculations
re-establish the double-peaked structure seen in the experimental data,
which was missing in the linear coupling calculations.
Indeed,
the barrier distribution obtained by
including the coupling to the octupole vibration of $^{16}$O
to all orders, looks very similar to that obtained by totally ignoring it,
apart from a shift in energy.  A shift of 2 MeV (dashed line) of the former
is required for the two calculated distributions to coincide.
This shift is consistent with the general conclusion that the main effect of
the coupling to an inelastic channel whose excitation energy is larger
than the curvature of the bare fusion barrier, {\it i.e.} an adiabatic
coupling, is to introduce a static potential shift~\cite{THAB94}. Thus
the shape of the barrier distribution does not change
unless the coupling is very strong and the coupling form factor
itself has a strong radial dependence.
The effects of these excitations can then be included in the `bare' 
potential in the coupled--channels calculations. Thus for
$^{16}$O~+~$^{144}$Sm, where potential parameters for the
calculations are obtained from
a fit to the high energy data, the effects of the octupole vibration of
$^{16}$O are already included. The explicit inclusion of the coupling to
octupole vibration then leads to double counting which manifests itself
by introducing an additional shift in the barrier (or barrier distribution)
as observed.

\section{APPLICATIONS}

The improved understanding of the fusion process as a result of the 
new measurements can lend itself to applications in other fields, {\it 
e.g.} population of high spin states for nuclear spectroscopy studies and
formation of super-heavy elements. These aspects were discussed in the 
previous talk by N. Rowley and  
here we will limit ourselves to a discussion of
the anomalous fission fragment
anisotropies, which  were explained in terms of our  
knowledge of the fusion barrier 
distributions.

\subsection{Nuclear orientation effects in fission}

The angular distribution for the fission process is characterised by the 
anisotropy, defined as the ratio of yield at 180$^\circ$ 
to that at 90$^\circ$ with respect to the beam axis 
W(180$^\circ$)/W(90$^\circ$). 
The anisotropies generally decrease with decreasing bombarding energies.
However, it was observed~\cite{opbdiff,largeans2}
that for reactions with actinide targets
at near-barrier energies, the anisotropies were not only
much larger than the  predictions of the 
transition state model~(TSM)~\cite{tsm}, but increased
as the beam energy dropped through the fusion barrier region. This 
observation agrees with the recently measured~\cite{hi95} anisotropies for the
$^{16}$O~+~$^{238}$U shown in the lower panel of Fig.~\ref{ou}.
The expected energy dependence of the anisotropy
based on the TSM 
calculation is indicated by the dot--dashed line. No
reasonable variation of parameters can reproduce the data.
It is clear that the observed anisotropies 
cannot be described by this model.

There exists a striking correlation between anisotropy and the 
barrier distribution;
 the low barriers are correlated with the higher anisotropies and the 
transition from higher to lower anisotropies occur at an energy close to the 
main peak. This observation prompted~\cite{hi95} a 
simple geometrical model to be proposed 
which is able to explain the data in a simple and
elegant way.
Considering the collision classically, the higher fusion barriers correspond
to contact of the projectile with the flattened side of the prolate
target, resulting initially in a compact dinuclear system.
Conversely,
the lower barriers correspond to contact with the tip, giving an
elongated dinuclear system. Intuitively, it seems reasonable that
the former configuration would be more likely to result in fusion-fission,
and the latter in quasi--fission (fission--like fragments from a system 
which never formed a compact compound nucleus inside the unconditional 
fission barrier).

It was assumed that collisions with the tips of the target nuclei
result in quasi-fission only. The assumed energy dependence of the
anisotropies for quasi--fission is shown by the dotted line in 
Fig~\ref{ou}.
The anisotropy for fusion-fission is taken from the transition
state model, and is shown by the dot-dashed line in the figure.
The average anisotropy, shown by the solid line in the figure, 
 was determined by taking the anisotropy assumed for
quasi-fission, and the fusion-fission anisotropies
calculated with the transition state model, and weighting each of them 
appropriately with a geometrical factor~\cite{hiprev}.
The calculations show that as we go from lower to higher 
energies there is  a gradual transition from anisotropies due to 
quasi--fission to those due to a mixture of quasi--fission and 
fusion--fission. The calculations follow the
trends of the data very well, resolving the observed anomaly and 
supporting the hypothesis that collisions
with the tips of the deformed target nuclei
result in quasi-fission, whilst
collisions with the sides result in fusion, followed almost always
by fission. 
This mechanism may have significance in 
explaining the failure of reactions with actinide targets to give 
larger cross--sections for the formation of very 
heavy nuclei than cold fission reactions with targets near Pb. 

\section{Summary} 

High precision measurements of fusion excitation functions 
have challenged the view that fusion cross-sections are smooth functions of 
energy which can be reproduced by any model which include some couplings. 
These  measurements, which enable the extraction of a representation 
of the barriers encountered during the fusion process,  provide 
an identification not only of the main couplings, but also the 
way in which they affect fusion. With this improved understanding 
of the fusion process we now may 
predict the dominant couplings likely to be present 
in a given reaction, benefiting fields which use fusion as a tool. 
The study of fusion reactions close to the 
barrier investigates features of coupling--assisted 
tunnelling, a phenomenon appearing in a diverse range of physical systems. It 
would 
be interesting to see whether one can exploit the ability in nuclear physics 
to vary the coupling strength and learn more about such problems.

% ========REFERENCES
\bibliographystyle{unsrt}

\newpage

\begin{figure}[h,t,b]
\caption{
Calculated excitation functions and barrier distributions for
a single-barrier (dashed lines) compared with three
coupling schemes involving coupling to a negative $Q$-value channel (a)
and (d), a positive $Q$-value channel (b) and (e), and coupling
associated with a nucleus with a permanent quadrupole deformation (c) and 
(f).
The cross--sections, 
plotted against the ratio of the energy to the average barrier $B_0$, are
essentially identical at low and high energies. The type of coupling is more
easily seen in the lower part of the figure than in the excitation functions
themselves. (Adapted from Ref.~9). }
\label{fig1}
\end{figure}

\begin{figure}[t,b]
\caption{Precise excitation functions are necessary to obtain meaningful 
barrier distributions. Two barrier distribution are shown, 
extracted from data with uncertainties of (a) 10\%  and 
(b) 1\%. The data in panel (a) is unable to 
distinguish between the two calculations shown by the dashed and solid 
lines.}
\label{fig2}
\end{figure}

\begin{figure}[t,b]
\caption{Comparison of the measured fusion excitation functions and 
the extracted barrier distributions for the indicated reactions. The
dotted curves are the result of single-barrier penetration calculations.
The dashed lines show the calculations with the inclusion of 
the quadrupole deformations of $^{154}$Sm and $^{186}$W; the  
solid lines include hexadecapole 
deformations. The  
differences in the shape of the barrier distribution is essentially due to 
the different sign of the hexadecapole deformations for the two nuclei.
(From Ref.~9).}
\label{deform}
\end{figure}

\begin{figure}[t,b]
\caption{
Comparison of the measured excitation functions (a and c) and extracted
barrier distributions (b and d) for $^{16}$O~+~$^{144}$Sm and for $^{17}$O~+~$^{144}$Sm. 
Couplings to the 2$^+$
and 3$^-$ states of $^{144}$Sm is adequate to describe the  $^{16}$O data as 
shown by the solid lines in the left panels. These couplings are however not 
sufficient for the $^{17}$O data (dashed lines in the right panels), which 
requires additional coupling to  a  neutron transfer channel with positive 
$Q$-value (solid lines). 
}
\label{sm144}
\end{figure}

\begin{figure}[t,b]
\caption{
Comparison of the experimental data with
calculations including couplings to one--phonon states  (left 
panels) and up to two--phonon states in $^{208}$Pb (right panels).
}
\label{opb}
\end{figure}

\begin{figure}[t,b]
\caption{Experimental 
fusion barrier distributions 
for the $^{40}$Ca~+~$^{192}$Os, $^{194}$Pt reactions; the arrow indicates the 
feature arising as a result of  coupling to the 3$^-$ state in $^{40}$Ca. 
Data are from Ref.~24.
}
\label{caos}
\end{figure}

\begin{figure}[t,b]
\caption{Fusion excitation function 
and the barrier distribution for the
$^{16}$O + $^{144}$Sm system compared with realistic 
coupled--channels calculations with the 
 linear coupling approximation (left panels) and all order coupling (right 
panels).  The solid lines are  the results when $^{16}$O is treated as inert.
The dashed lines are the results when
the coupling to the octupole vibration of $^{16}$O is also
taken into account; the long--dashed line in the right panel 
is the same calculation as the dashed
line, with but with the average barrier increased by 2 MeV.  
 Effects of the octupole vibration of
$^{144}$Sm are taken into account in all the calculations. (Adapted from 
Ref.~28).}
\label{allord}
\end{figure}

\begin{figure}[t,b]
\caption{
Panel (a) shows the experimental 
fusion barrier distribution for  the $^{16}$O + $^{238}$U reaction.
The effects of static deformation of $^{238}$U have been included in the 
calculations shown by the solid line. 
The fission anisotropies are shown in panel (b).
The expected anisotropy based on the statistical model
calculation is indicated by the dot-dashed line, whilst
the assumed quasi-fission anisotropy is given by the dotted line.
The model described in the text gives the solid curve. 
}
\label{ou}
\end{figure}

\end{document}